\documentclass[openacc]{rsproca_new}

\begin{document}

\title{Supermassive Black Holes in the Early Universe}

\author{F. Melia$^{1}$ and T. M. McClintock$^{2}$}

\address{$^{1}$Department of Physics, The Applied Math Program, and Department of Astronomy,
The University of Arizona, AZ 85721, USA\\
$^{2}$Department of Physics, The University of Arizona, AZ 85721, USA}

\subject{xxxxx, xxxxx, xxxx}

\keywords{xxxx, xxxx, xxxx}

\corres{F. Melia\\
\email{fmelia@email.arizona.edu}}

\begin{abstract}
The recent discovery of the ultraluminous quasar SDSS J010013.02+280225.8 at
redshift 6.3 has exacerbated the time compression problem implied by the appearance
of supermassive black holes only $\sim 900$ Myr after the big bang, and only $\sim 
500$ Myr beyond the formation of Pop II and III stars. Aside from heralding the
onset of cosmic reionization, these first and second generation stars could have
reasonably produced the $\sim 5-20\;M_\odot$ seeds that eventually grew into $z\sim 
6-7$ quasars. But this process would have taken $\sim 900$ Myr, a timeline that
appears to be at odds with the predictions of $\Lambda$CDM without an anomalously
high accretion rate, or some exotic creation of $\sim 10^5\;M_\odot$ seeds. There
is no evidence of either of these happening in the local universe. In this paper,
we show that a much simpler, more elegant solution to the supermassive
black hole anomaly is instead to view this process using the age-redshift relation
predicted by the $R_{\rm h}=ct$ Universe, an FRW cosmology with zero active mass.
In this context, cosmic reionization lasted from $t\sim 883$ Myr to $\sim 2$ Gyr
($6\lesssim z\lesssim 15$), so $\sim 5-20\;M_\odot$ black hole seeds formed
shortly after reionization had begun, would have evolved into $\sim 10^{10}\;
M_\odot$ quasars by $z\sim 6-7$ simply via the standard Eddington-limited
accretion rate. The consistency of these observations with the age-redshift
relationship predicted by $R_{\rm h}=ct$ supports the existence of dark
energy; but not in the form of a cosmological constant.
\end{abstract}


\begin{fmtext}
\section{Introduction}
The recent discovery of SDSS J010013.02+280225.8 (henceforth J0100+2802), an
ultraluminous quasar at redshift $z=6.30$, has accentuated the problem of
supermassive black-hole growth and evolution in the 
\end{fmtext}

\maketitle

\noindent early Universe
\cite{1}. Each of the $\sim 50$ previously discovered quasars
at redshifts $z>6$ \cite{2,3,4,5,6,7,8,9} contains a black hole
with mass $\sim 10^9\;M_\odot$, challenging the standard model's predicted
timeline, which would have afforded them fewer than $900$ Myr to grow after
the big bang, but likely even fewer than $\sim 500$ Myr since the onset of
Population II star formation. With an estimated mass of $\sim 10-12 \times 
10^{9} \;M_\odot$, ten times greater than the others, J0100+2802
significantly compounds this time-compression problem.

The early appearance of supermassive black holes is an enduring
mystery in astronomy. Such large aggregates of mass could not
have formed so quickly in $\Lambda$CDM without some anomalously
high accretion rate \cite{10} and/or the creation of unusually
massive seeds \cite{11}, neither of which has actually ever
been observed. In the local Universe, black-hole seeds
are produced in supernova explosions, which sometimes leave behind
$\sim 5-20\; M_\odot$ remnant cores, far from the $\sim 10^5\;
M_\odot$ objects required to grow via Eddington-limited accretion
into the billion solar-mass quasars seen at redshifts $\sim 6-7$.
Any attempt at circumventing this problem is severely challenged
by the fact that no high-$z$ quasar has ever been observed to accrete
at more than $\sim 1-2$ times the Eddington rate (see, e.g., figure~5
in Ref.~\cite{12}).

In this paper, however, we demonstrate that such exotic processes
are actually not necessary to address this apparent anomaly; although the timeline
implied for the early Universe by the existence of J0100+2802 and its
brethren may be problematic in $\Lambda$CDM, it is fully consistent
with standard astrophysical processes in the $R_{\rm h}=ct$ Universe
\cite{13,14}, an FRW cosmology with zero active mass \cite{15}.
Indeed, this tantalizing discovery comes on the heels of recent high-precision
Baryon Acoustic Oscillation (BAO) measurements \cite{16,17,18},
that have apparently ruled out the concordance model in favor of
$R_{\rm h}=ct$ at better than the 99.95\% C.L. \cite{19}.
A resolution of the time-compression problem implied by J0100+2802
therefore provides important and timely observational confirmation
of these critical BAO results.

\section{Black Hole Growth in the Early Universe}
In the standard picture, the Universe became transparent about
0.4 Myr after the big bang, descending into darkness as the
thermal radiation field shifted towards longer wavelengths.
The so-called Dark Ages ended several hundred Myr later, when
density perturbations condensed into stars and early galaxies,
producing ionizing radiation. Current constraints \cite{20}
place the epoch of re-ionization (EoR) at $z\sim 6-15$ which,
in the context of $\Lambda$CDM, corresponds to a cosmic time
$t\sim 400-900$ Myr (see figure~1).

\begin{table*}
\begin{center}
{\tiny
\caption{Highest-redshift Quasars}
\begin{tabular}{lllcrrccl}
&&& \\
\hline\hline
&&&&&&&& \\
Source & $\;\;\;$Redshift & $\;\,$FWHM & $F_{3000\AA}$ &  M$_{R_{\rm h}=ct}$ & M$_{\Lambda{\rm CDM}}$ &
Age$_{R_{\rm h}=ct}$ & Age$_{\Lambda{\rm CDM}}$ & Ref. \\
&& $\;\,$(km/s)& ($10^{-29}$ erg/cm$^{2}$/s/Hz)&($10^9M_{\odot}$)&($10^9M_{\odot}$)& (Gyr)& (Gyr) &\\
&&&&&&&& \\
\hline\hline
J0100+2802 & 6.300$\pm{0.010}$ & 5130$\pm{ 150}$ & 70.62$\pm{10.54}$ & 10.76$\pm{1.02}$ & 10.40$\pm{0.99}$ & 1.99 & 0.87 &
\cite{1} \\
P167-13& 6.508$\pm{0.001}$ & 2350$\pm{ 470}$ & 2.85$\pm{1.66}$ & 0.36$\pm{0.18}$ & 0.35$\pm{0.17}$ & 1.94 & 0.84 &
\cite{21} \\
P036+03& 6.527$\pm{0.002}$ & 3500$\pm{ 875}$ & 8.79$\pm{6.52}$ & 1.43$\pm{0.89}$ & 1.38$\pm{0.86}$ & 1.93 & 0.84 &
\cite{21} \\
J0305-3150& 6.604$\pm{0.008}$ & 3189$\pm{  85}$ & 3.18$\pm{0.04}$ & 0.87$\pm{0.05}$ & 0.84$\pm{0.05}$ & 1.91 & 0.82 &
\cite{21} \\
P338+29& 6.658$\pm{0.007}$ & 6800$\pm{1050}$ & 2.21$\pm{1.03}$ & 2.95$\pm{1.14}$ & 2.83$\pm{1.09}$ & 1.90 & 0.81 &
\cite{21} \\
J0109-3047& 6.745$\pm{0.009}$ & 4389$\pm{ 380}$ & 1.95$\pm{0.26}$ & 1.25$\pm{0.23}$ & 1.20$\pm{0.22}$ & 1.88 & 0.80 &
\cite{21} \\
J2348-3054& 6.886$\pm{0.009}$ & 5446$\pm{ 470}$ & 1.64$\pm{0.57}$ & 1.84$\pm{0.45}$ & 1.75$\pm{0.43}$ & 1.84 & 0.78 &
\cite{21} \\
J1120+0641& 7.085$\pm{0.003}$ & 3800$\pm{ 200}$ & 6.39$\pm{0.49}$ & 1.36$\pm{0.15}$ & 1.29$\pm{0.14}$ & 1.80 & 0.75 &
\cite{7} \\
\hline\hline
\end{tabular}
}
\end{center}
\end{table*}

The best probes of the re-ionization process are actually the
high-$z$ quasars themselves. The absence of structure bluewards
of their Ly-$\alpha$ restframe emission observed by SDSS at $z\gtrsim 6$
suggests a decreasing ionizing fraction along the line-of-sight
\cite{22}. That the neutral fraction approaches $\sim 1$
by $z\sim 15$ is supported by the Wilkinson Microwave Anisotropy
Probe (WMAP \cite{23}) mission, whose measurements show
that the Universe was $\sim 50\%$ neutral at $z\gtrsim 10$, with
re-ionization starting before $z\sim 14$ \cite{24}.

\begin{figure}[!h]
\centering\includegraphics[width=5.0in]{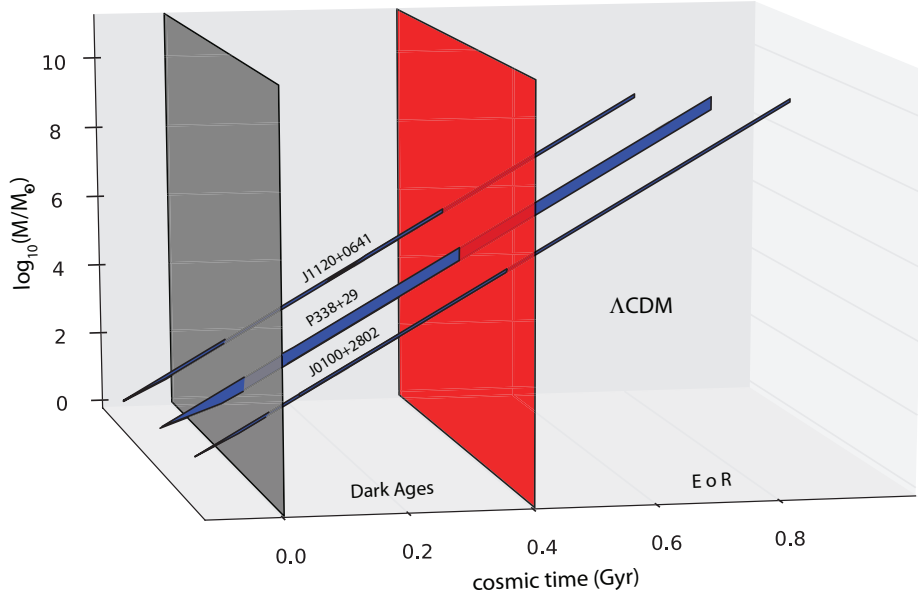}
\caption{Growth of quasars J0100+2802,
P338+29, and J1120+0641, versus $t$ in $\Lambda$CDM, with
concordance parameter values: $\Omega_{\rm m}=0.31$, $k=0$, $w_\Lambda=-1$ and
$H_0=67.3$ km s$^{-1}$ Mpc$^{-1}$. The range in timelines
corresponds to $M_{\rm seed}=5-20\;M_\odot$. The Dark Ages evolved into
the EoR at $t\sim 400$ Myr ($z\sim 15$), which lasted until
$\sim 900$ Myr ($z\sim 6$). For $M_{\rm seed}=20\;M_\odot$ and
Eddington-limited accretion, these three black holes would have had masses
$3.0\times 10^5\;M_\odot$, $3.1\times 10^5\;M_\odot$, and $5.4\times 
10^5\;M_\odot$, respectively, at the start of the EoR. The seeds would have
been created $\sim 33$ Myr, $\sim 35$ Myr, and $\sim 59$ Myr
{\em prior} to the big bang.}
\label{figure1}
\end{figure}

Standard astrophysical principles suggest that ionizing radiation
was produced by Pop II and III stars. More exotic physics, invoking
the decay of dark-matter particles or cosmic strings, is poorly
known and, anyway, appears to be too tightly constrained to account
for the EoR on its own \cite{25}. Almost certainly,
high-$z$ quasars became more important towards the end of the EoR,
though they probably could not have been the dominant source of
ionizing radiation for the whole EoR \cite{26}.

Our view of how the Universe evolved through the dark ages and
into the EoR is informed by many detailed simulations that have
been carried out in recent years \cite{27,28,29,30,31,32,33,34,35,36}.
(For recent reviews, see Refs.~\cite{37,38}.)
According to these calculations, the first (Pop III) stars formed by redshift
$\sim 20$ at the core of mini halos with mass $\sim 10^6\;M_\odot$
\cite{39,40,41,42}. In the concordance
$\Lambda$CDM model with parameter values $\Omega_{\rm m}=0.31$,
$k=0$, $w_\Lambda=-1$ and Hubble constant $H_0=67.3$ km s$^{-1}$
Mpc$^{-1}$, this redshift corresponds to a cosmic time $t\sim 200$ Myr.

Based on the astrophysics we know today, specifically the rate of
cooling in the primordial gas, it is difficult to see how Pop III
stars could have formed earlier than this. The subsequent transition
to Pop II star formation incurred additional delays because the
gas expelled by the first generation of Pop III stars had to cool and
re-collapse. Detailed calculations of this process show that the
gas re-incorporation time was at least $\sim 100$ Myr \cite{43,44}.

So the formation of structure could not have begun in earnest until
at least $\sim 300$ Myr after the big bang, and the start of
the EoR presumably overlapped with the ramp up in Pop II star
formation and evolution. Standard astrophysical principles would
suggest that this was also the time when $\sim 5-20\;M_\odot$
black-hole seeds were created, presumably following the supernova
explosion of {\em evolved} Pop II (and possibly Pop III) stars.
In other words, unless we introduce new, exotic physics for the
formation of these seeds, it is difficult to see how they could
have emerged any earlier than the start of the EoR, and certainly
no sooner than the transition from Pop III to Pop II star formation.

Without invoking anomalously high accretion rates, these seeds
would have grown at a rate set by the Eddington limit, defined
to be the maximum luminosity attainable from the outward radiation
pressure on ionized material under the influence of gravity. For
hydrogen plasma, this power is given as $L_{\rm Edd}\approx
1.3\times 10^{38}(M/M_\odot)$ ergs s$^{-1}$, in terms of the
black-hole mass $M$. The mass accretion rate $\dot{M}$ is inferred
from $L_{\rm Edd}$ with the inclusion of an efficiency $\epsilon$ for
converting rest-mass energy into radiation. A minimum of $\sim
6\%$ is expected for $\epsilon$ in a Schwarzschild black hole, but
other factors (such as a black-hole spin) may enhance $\epsilon$
above this value. To allow for such variations, one typically
adopts the fiducial value $\epsilon=0.1$.

Conventional astrophysics would then suggest that early black-hole
growth was driven by an accretion rate $\dot{M}=L_{\rm Edd}/\epsilon c^2$,
and solving for the mass as a function of time one then arrives
at the so-called Salpeter relation,
\begin{equation}\label{2.1}
M(t) = M_{\rm seed}\exp\left({t-t_{\rm seed}\over 45\;{\rm Myr}}\right),
\end{equation}
where $M_{\rm seed}$ ($\sim 5-20\;M_\odot$) is the seed mass produced
at time $t_{\rm seed}$. It is now straightforward to see why the
discovery of J0100+2802 at redshift 6.3 presents such a big problem
for the standard model. With an inferred mass of $\sim 10-12\times
10^9\;M_\odot$, the minimal growth time implied by the Salpeter
relation is $t-t_{\rm seed}\sim 910$ Myr (assuming conservatively
that $M_{\rm seed} = 20\;M_\odot$). And since in the standard model
$t(z=6.3)\sim 880$ Myr,
not only is this quasar inconsistent with what is known about the
transition from the Dark Ages to the EoR, but it would have had to
start growing {\it before} the big bang, an obviously non-sensical
interpretation.

Of course, these are the reasons some workers have been driven to
find non-standard physics to account for the discrepancy between this
implied timeline and the predictions of the standard model. But given that
no evidence has been seen for such proposals, the only viable explanation
within the context of $\Lambda$CDM appears to be the possible role played by
mergers in the early Universe \cite{45,46,47}.
However, even this mechanism could only work if the black-hole seeds formed
well before the EoR, and their creation would have had to end by $z\sim 20-30$
to avoid over-producing the low-mass end of the distribution. In other words,
the black-hole seeds would have had to be produced by unknown, exotic
processes unrelated to the Pop II and Pop III star formation rate. Also,
a more recent consideration of possible mergers \cite{48} suggests
that such events could not have been common in the early Universe. Simulations
now show that Pop III stars needed very large halos to form, which would
have decreased the halo abundance by orders of magnitude. In addition,
the associated build-up of Lyman-Werner background radiation would have
accelerated the dissociation of H$_2$, which was necessary for the star-formation
process. Both of these trends point to a reduction in the Pop III star formation
rate, and therefore a reduction in the possible number of seed black holes.
All in all, the mysterious appearance of billion-solar mass quasars at $z\sim 6-7$
therefore constitutes significant tension with the standard ($\Lambda$CDM) model.

\section{A Cosmological Solution}
The purpose of this paper is to demonstrate that a resolution of the
time-compression problem revealed by supermassive quasars at high redshifts
may be found more reasonably in the cosmology itself, rather than the
physics of black-hole birth and evolution. Recent observations have pointed
to a growing inconsistency with $\Lambda$CDM \cite{19}, so it would not
be surprising to find that its predicted age-redshift relation is also at
odds with the quasar measurements. Table 1 lists the highest
redshift quasars discovered to date, including the most massive (J0100)
among them. The growth histories of three representative members of this
group calculated using the Salpeter relation (equation~1) are illustrated
in figure~1, which also shows the duration of the Dark Ages and EoR.

The inferred black-hole mass $M$ in this table is calculated from the simultaneous
measurement of the quasar's luminosity and the velocity of its line-emitting gas via the
observation of its Doppler-broadened Mg II line \cite{49}. This is made possible
by reverberation mapping, which produces a tight relationship between the distance of the
line-emitting gas from the central ionizing source, and the optical/UV luminosity
\cite{50}. When high-quality line and continuum measurements are available,
one can infer the black-hole mass from the relation \cite{51}
\begin{equation}\label{3.1}
\log M = 6.86 + 2\log{{\rm FWHM(MgII)}\over 1,000\;{\rm km}\;{\rm s}^{-1}}+ 
0.5\log{L_{3000\;\AA}\over 10^{44}\;{\rm ergs}\;{\rm s}^{-1}}\;,
\end{equation}
in which the luminosity $L_{3000\;\AA}$ at rest-frame wavelength $3000\;\AA$ is 
calculated from the measured flux density $F_{3000\;\AA}$ separately for each 
assumed cosmology, which determines the luminosity distance. When needed, the 
bolometric luminosity is determined from $L_{3000\;\AA}$ using a bolometric 
correction factor $\eta$. The process of identifying $\eta$ for these sources 
is somewhat tricky, but reliable \cite{52}, yielding the now commonly used value 
$\eta\sim 6.0$ \cite{53}. The flux density $F_{3000\;\AA}$ is measurable to an 
accuracy of about $10\%$, while the FWHM is accurate to about $15\%$. Equation~(2) 
therefore yields mass estimates accurate to about a factor $\sim 2$ (i.e., 
$0.4-0.5$ dex) in most cases.

Of course, the caveat with the use of Equation~(3.1) for this purpose is 
that the numerical scale $6.86$ was obtained using several thousand high-quality
spectra from the SDSS DR3 quasar sample \cite{54,55}, with a calibration to
the H$\beta$ and C IV relations using the luminosity distance in the concordance
model. The fact that $L_{3000}$ appears in this expression means that one
ought to separately recalibrate these values for each different cosmology,
because the luminosity distance $d_L^{R_{\rm h}=ct}$ in $R_{\rm h}=ct$ is not 
the same as that ($d_L^{\Lambda{\rm CDM}}$) in $\Lambda$CDM. However, it is 
well known by now that the ratio of these distances is very close to unity 
all the way to $z>6-7$ (see, e.g., Figure~3 in Ref.~\cite{56}). For example,  
$0.9\lesssim d_L^{\Lambda{\rm CDM}}/d_L^{R_{\rm h}=ct}\lesssim 1.1$ over
the redshift range $4\lesssim z\lesssim 8$. Thus, the error implied for
$M$ in Equation~(3.1) due to differences in calibration between the two
models is $\lesssim 5\%$, which is well within the overall expected
error of a factor $2$ in this expression. For the purpose of this paper,
it is therefore safe to ignore such differences in calibration between
$R_{\rm h}=ct$ and $\Lambda$CDM.

In the $R_{\rm  h}=ct$ Universe, the age-redshift relation is simply 
given by the expression
\begin{equation}\label{3.2}
1+z = {1\over H_0t}
\end{equation}
where, for ease of comparison between the two models, we use the
same value $H_0=67.3$ km s$^{-1}$ Mpc$^{-1}$ in both cases. In 
this model, the dark ages therefore ended---and the EoR began---at
approximately $883$ Myr (i.e., $z\sim 15$), and reionization was
completed by about 2 Gyr ($z\sim 6$). By comparison, the EoR lasted
from $\sim 400$ Myr to $\sim 900$ Myr in $\Lambda$CDM. Insofar as
the EoR is concerned, what we do know fairly well from observations
is its redshift range, but the corresponding ages are less reliably 
known. However, there are several indications that the time compression 
problem associated with supermassive black holes occurs more generally. 
For example, the early appearance of galaxies at $z\sim 10-12$ may be
an even bigger problem than that for quasars \cite{57} because, whereas 
a single event could have created a quasar, galaxies had to form 
gradually through the assembly of more than $10^9$ stars. This does
not appear to be feasible in such a short time following the transition
from Pop III to Pop II stars with the $2-4\;M_\odot$ yr$^{-1}$
star-formation rate predicted in the standard model. 

\begin{figure}[!h]
\centering\includegraphics[width=5.0in]{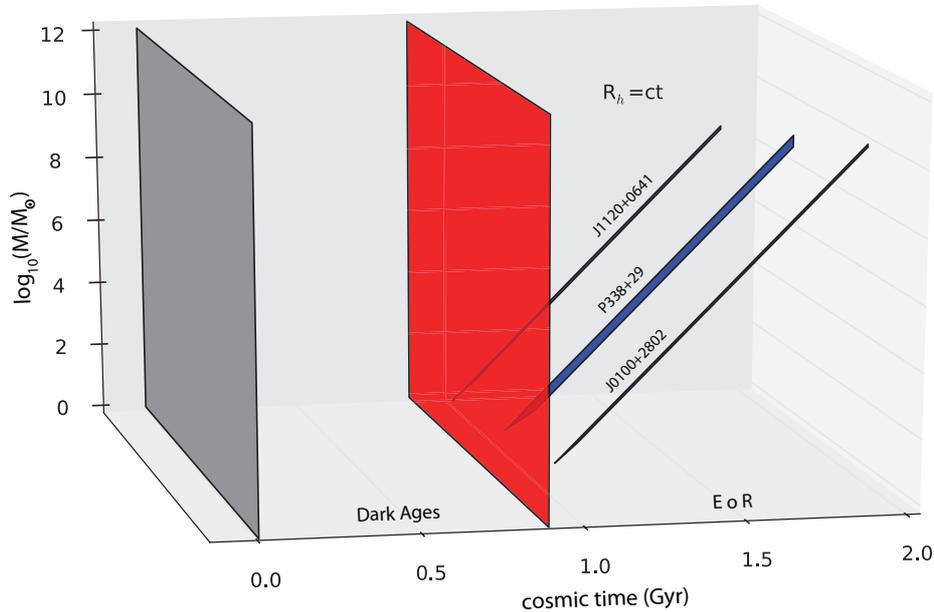}
\caption{Same as figure~1, but now for the $R_{\rm h}=ct$
cosmology, which has only one free parameter: $H_0$, whose value is
assumed to be the same as in $\Lambda$CDM for ease of comparison. Here, the
EoR lasted from $\sim 883$ Myr to $\sim 2.0$ Gyr ($6\lesssim z\lesssim 15$),
and the $20\;M_\odot$ seeds would have formed at $t\sim 1.09$ Gyr (J0100+2802),
$\sim 1.06$ Gyr (P338+29), and $\sim 990$ Myr (J1120+0641), shortly after
the start of the EoR.}
\label{figure2}
\end{figure}

This question of how early galaxies formed is probably also linked to 
the source of ionizing radiation in the intergalactic medium. With the
timeline afforded them in $\Lambda$CDM, the known sources of UV ionizing 
radiation, principally galaxies at intermediate redshifts and dwarf 
galaxies at higher redshifts, may have produced most of the ionizing 
radiation leaking into the intergalactic medium, but only for an inferred 
escape fraction of about $20\%$, which is not consistent with other 
measurements that suggest a value as small as $\sim 5\%$. However, if the
EoR is viewed in the context of $R_{\rm h}=ct$, the extended timeline
reduces the required escape fration to a value much closer to 
$\sim 5\%$ \cite{58}.

A comparison between the quasar timelines in figures~1 and 2 highlights the 
significant differences between these two models when it comes to how and 
when these supermassive black holes were formed. In $R_{\rm h}=ct$, all of 
the quasars seen at $z\sim 6-7$ could have easily grown to their measured size
via Eddington-limited accretion within the EoR. Crucially, all of the required
$\sim 5-20\;M_\odot$ seeds would have formed {\em after} the end of the Dark
Ages, presumably from the supernova explosion of evolved Pop II and/or III stars.
The birth and growth of these high-$z$ quasars is therefore entirely consistent
with standard astrophysical processes and our current understanding (\S~II) of
how the first stars formed and evolved into nascent structure in the early Universe.

\section{Conclusions}
Irrespective of whether or not mergers played a role in building up the black-hole
mass distribution, black-hole seeds in $\Lambda$CDM would have had to form very
early in the Universe's history, well before the onset of re-ionization at $z\sim 15$.
Unless some unknown exotic mechanism was responsible for their creation, they
would almost certainly have been produced by Pop III stars shortly after the big
bang. However, no evidence of re-ionization has yet been found prior to the redshift
($z\sim 20-30$) where seed creation must have stopped in order to avoid overproducing
lower mass black holes. Reconciling these opposing trends may not be feasible.

In this paper we have assumed that the high-$z$ quasars accreted steadily
at the Eddington rate. Their duty cycle is unknown, however, so it's possible that,
on average, their rate of growth was less than Eddington. This would lengthen
the Salpeter time ($\sim 45$ Myr) even further, thereby exacerbating the time compression
problem for $\Lambda$CDM. On the other hand, circumstantial evidence does
suggest that when they were turned on, high-$z$ quasars did accrete at close
to Eddington. This is based on the maximum black-hole mass observed
in the local Universe \cite{59}. No more than a couple of $10^{10}\;M_\odot$
black hole masses have thus far been detected, even after the peak quasar
activity at $1\lesssim z\lesssim 3$ \cite{60}. Yet high-$z$ quasars
would have had to be more massive than $10^{10}\;M_\odot$ in order to
produce their fluxes measured at Earth if they were accreting below the
Eddington limit.

Perhaps the solution to the time-compression problem in $\Lambda$CDM
is simply that the high-$z$ quasars accreted steadily at super-Eddington
rates from the time their seeds formed all the way to when we see them
at $\sim 900$ Myr. But one can see from Equation~(2.1) that in order to 
grow to $\sim 10^{10}\;M_\odot$ in $\sim 400-500$ Myr, they would have
had to accrete at $\sim 2-3$ times the Eddington rate throughout that
epoch. One cannot yet completely discount such a possibility, but
we should then be able to find at least some members of this super-Eddington
class at $z\gtrsim 6$. However, all the current observations appear to
have ruled out the existence of such sources. The latest measurements
\cite{61,62,63} indicate that the most distant quasars are accreting 
at no more than the standard Eddington value, and their accretion rate
is decreasing at smaller redshifts.

When all the facets of the time-compression problem are
considered together, the simplest and most elegant solution to the early
appearance of supermassive black holes appears to be a change in the
cosmology itself.  In the $R_{\rm h}=ct$ Universe, the birth, growth and
evolution of high-$z$ quasars are fully consistent with the principal timescales
associated with Pop II and III star formation, and the ensuing epoch of
reionization. This picture supports the view that supermassive black
holes probably did not contribute significantly to the ionizing radiation
early on, but may have become more prominent contributors towards
the end of the EoR. Indeed, it may turn out that the rapid ramp up in
black-hole mass towards $z\sim 6$ (see figure~2) may have been
responsible for completely ionizing the intergalactic medium, thereby
bringing an end to the EoR around $z\sim 6$.

Our conclusion adds some support to the possibility, already suggested
by other kinds of observation---such as the latest high-precision Baryon
Acoustic Oscillation data \cite{16,17,19}---that
$\Lambda$CDM does not account for high-precision cosmological measurements
as well as $R_{\rm h}=ct$ does. The latter is itself an FRW cosmology, though
restricted by the zero active mass condition (i.e., $\rho+3p=0$), which is
lacking in $\Lambda$CDM \cite{15}. With this constraint, however,
dark energy cannot be a cosmological constant. Its density must evolve
dynamically, suggesting an origin in particle physics beyond the standard
model. The analysis we have reported in this paper therefore has significant
implications because, unlike many other kinds of cosmological measurements,
the early appearance of quasars hinges on the time-redshift relationship
rather than on integrated distances. It has the potential of opening up
a new perspective on the expansion histories in different models in that
crucial early period ($\lesssim 1-2$ Gyr) when the dynamics differed
significantly from one cosmology to the next.

\section*{Acknowledgments}
FM is grateful to Purple Mountain Observatory in Nanjing, China, where some
of this work was carried out. He acknowledges partial support from the Chinese 
Academy of Sciences President's International Fellowship Initiative, under 
Grant No.\ 2012T1J0011.

\vskip 0.2in\noindent{\bf Ethics Statement.} This research poses no ethical considerations.

\vskip 0.2in\noindent{\bf Data Accessibility Statement.} All data used in this paper were
previously published, as indicated in Table 1.

\vskip 0.2in\noindent{\bf Competing Interests Statement.} We have no competing interests.

\vskip 0.2in\noindent{\bf Authors' contributions.} FM conceived the project and wrote the
paper. TMM carried out the simulations. All authors gave final approval for
publication.

\vskip 0.2in\noindent{\bf Funding.} The Chinese Academy of Sciences President's International
Fellowship Initiative, Grant No.\ 2012T1J0011.

\end{document}